\documentclass{article}
\usepackage{frascatiphys_R}
\newcommand{\etacprime}{\eta_c'}
\newcommand{\mev}{\,\mbox{MeV}}
\newcommand{\gev}{\,\mbox{GeV}}
\newcommand{\kev}{\,\mbox{keV}}
\newcommand{\nb}{\,\mbox{nb}}
\newcommand{\pdp}{\psi(3770)}
\newcommand{\pp}{\psi(2S)}
\newcommand{\jpsi}{J/\psi}
\newcommand{\bbbar}{b \bar b}
\newcommand{\ccbar}{c \bar c}
\newcommand{\ppbar}{p \bar p}
\newcommand{\qqbar}{q \bar q}
\newcommand{\DDbar}{D \bar D}
\newcommand{\pizero}{\pi^0}
\newcommand{\chicj}{\chi_{cJ}}
\newcommand{\invfb}{\,\mbox{fb}^{-1}}
\newcommand{\invpb}{\,\mbox{pb}^{-1}}
\newcommand{\etal}{{\it et al.} }
\begin{document}
\title{ 
\quad \hfill PUZZLE PIECES: \hfill \quad
\newline
RESULTS ON \boldmath $\bbbar$ AND $\ccbar$ SPECTROSCOPY AND DECAY
}
\author{
Hanna Mahlke-Kr\"uger \\
\em Cornell University, Ithaca, NY 14853  
}
\maketitle
\baselineskip=11.6pt
\begin{abstract}
Recent results in the field of Heavy Quarkonia are reviewed,
with results either providing new precision measurements or
addressing key unanswered questions.
\end{abstract}
\baselineskip=14pt

\section{Introduction}
Heavy quarkonia exhibit features similar to the positronium spectrum:
a discrete system of states with the spacings and transition rates 
dictated
by the binding force, which in this case is the strong interaction. 
Investigating heavy quarkonia therefore enables us to study
important aspects of QCD.
While heavy quarkonia parton level decay by
annihilation is a perturbatively calculable process, transitions 
among them are not
as they are soft due to the energy spread between the states, which is
below $1\gev$. 

Theory has made progress recently that indicates the need for experimental
results at the few percent level in precision. On the other hand, there are 
important unanswered questions where experimental information is scant.
The following results have been selected so as to address one or the other.

In view of the very limited space available for this report, 
no figures are shown, but references
to publications where they can be found are given. More $\pp$~results from 
BES were presented in a separate talk by X.H. Mo at this conference.

\section{Spectroscopy}
\subsection{Measurements of the $\etacprime$ Mass\cite{etacprimeexp}}
After the first evidence for the $\etacprime$ more than twenty years ago, 
which established it 
from the direct M1 transition $\pp \to \gamma \etacprime$,
the experimental picture has consolidated in the past two years:
In $B \to \etacprime K$, $e^+e^- \to \jpsi \etacprime$, and
$\gamma\gamma \to \etacprime$ studies,
the $\etacprime$ mass is found to be around $3638\mev$, or
$44\mev$ higher than measured before. This means that the
$2^3S_1$-$2^1S_0$~mass splitting is reduced by a factor of
two, and is now two times smaller than the hyperfine splitting 
at $n=1$.  
Comparing these two is interesting because, due to the
difference in $\ccbar$ distance, they sample different
areas of the binding potential, which connects the confinement region
with that of asymptotic freedom.

\subsection{$X(3872)$\cite{x3872}}

Since the discovery of the ``$X(3872)$'' by Belle and 
subsequent confirmation by BaBar, CDF, and D0, several attempts 
to explain this narrow state have been made on the theory side. Among the
plausible ones are that it could be a charmonium state, a $D \bar D$~molecule,
or even an exotic state. Experimental efforts have focussed on
studying decay or production modes that can clarify the nature of
this state by virtue of establishing its quantum numbers. 
The decay mode $X \to \pi^+\pi^- \jpsi$, which gives rise to the
state's characterization as ``charmonium-like'', remains the only
one seen so far. The dipion mass distribution is of special interest
as one hopes to answer the question whether or not the decay proceeds
through an intermediate $\rho$. In this context, searching for $X(3872)
\to \pizero\pizero\jpsi$ is of special importance. 
CLEO has engaged in a search for X(3872) in two-photon fusion and 
ISR~production, using $15\invfb$ of data at $\sqrt s = 9.46 - 11.30\gev$.
This allows access to $J^{PC}=1^{--}$ and $2n^{\pm +}$.
Preliminary upper limits have been placed:
$\Gamma_{ee} \times {\cal B}(X \to \pi^+\pi^- \jpsi)<6.8\,\mbox{eV}$ or
1\% of the production rate of $\pp$ in ISR events (assuming
a similar branching fraction ${\cal B}_{\pi^+\pi^-\jpsi}$),
and $(2J+1)\Gamma_{\gamma \gamma} \times {\cal B}(X \to \pi^+\pi^-\jpsi)
<16.7\,\mbox{eV}$, or one tenth of the $\eta_c$ production rate in
two-photon fusion. A similar ISR study has been done of BES data, using 
using $22.3\invpb$
at $\sqrt s = 4.03\gev$, which arrived at an upper limit of 
$\Gamma_{ee} \times {\cal B}(X \to \pi^+\pi^- \jpsi)
<10\,\mbox{eV}$.

\subsection{Transitions\cite{cleo:omegatransition}}

Transitions between states of heavy onia are by emission of
photons or hadrons such charged pion pairs, neutral single pions or pion pairs, 
and etas. In bottomonium, also an $\omega$ transition 
has recently been observed
as the first non-pionic hadronic transition 
in $\Upsilon(3S) \to \gamma \chi_{b1,2}(2S)$, 
$\chi_{b1,2} \to \omega \Upsilon(1S)$. 
The branching fractions are found to be substantial and also in
compliance  with a prediction for them to be
about equal:
${\cal B}(\chi_{b1[2]} \to \omega\Upsilon(1S)=
(1.63^{+0.31}_{-0.32}\mbox{}^{+0.15}_{-0.11})$
$[(1.10^{+0.35}_{-0.28}\mbox{}^{+0.16}_{-0.10})]\%$. 
Radiative decays to $\Upsilon$s are, to date, the only other known 
exclusive decay mode of the $\chi_{bJ}$ states, and are only a factor 
5-6 more common.

While $\eta$ and single $\pizero$ transitions have been seen in charmonium,
with recent BES studies showing a much increased precision over previous
results, a similar 
measurement in bottomonium is yet to be made. 

Dipion transitions are the most common ones both in $\ccbar$ and $\bbbar$.
Naively, one would expect that the ratio of branching fractions for
neutral and charged modes would be, related by isospin, 1:2. 
A direct measurement of this quantity resulted in
${\cal B}(\pp \to \pi^0\pi^0\jpsi)/{\cal B}(\pp \to \pi^+\pi^-\jpsi) =  
0.570 \pm 0.009 \pm 0.026$;
taking the
most recent PDG values for the individual branching fractions yields 
$0.59 \pm 0.04$. 
An interesting new measurement has been made by BaBar, using radiative return
events to the $\pp$ in $90\invfb$ of $\Upsilon(4S)$ data.
They find ${\cal B}(\pp \to \pi^+\pi^- \jpsi) = 0.361 \pm 0.40$, which 
decreases the ratio by over $12\%$, thereby bringing it within reach of $0.5$.

\section{Decays}
\subsection{$\pdp \to $ non-$\DDbar$?\cite{psi3770nonddbar}}
The experimental indication for the existence of a significant
$\pdp$
non-$\DDbar$ hadronic decay width stems from the difference between 
early total hadronic and the $D$~pair production cross
section measurements: $\sigma (\pdp \to D\bar D) = 5.0 \pm 0.5\nb$, 
$\sigma (\pdp \to hadrons) = 7.8 \pm 0.8\nb$.
This invites the the following set of questions: Which
non-$\DDbar$ channels are available to $\pdp$ decay?
Can the measurement of the total hadronic cross section be
confirmed? Can the measurement of the $D$-pair production
cross section be confirmed?

As to the last question, preliminary measurements seem to
indicate a higher $D$-pair production cross section:
$\sigma(\pdp \to D\bar D)^{CLEO} = (5.78 \pm 0.11 \pm 0.38)\nb$,
$\sigma(\pdp \to D\bar D)^{BES} = (6.51 \pm 0.44 \pm 0.39)\nb$.
The experimental techniques are somewhat
different in that BES tags one of the $D$~mesons, thereby
gaining statistical advantage, while CLEO tags both $D$~mesons,
resulting in independence from external branching fractions.
While there is an indication that the gap might not be as
wide as previously thought, about 20\% of the total 
width of $(23.6 \pm 2.7)\mev$\cite{pdg2004}
remain currently unaccounted for.     
Convincing unanimous evidence for 
what this gap is filled by has yet to be presented. The BES
collaboration measured  ${\cal B}(\pdp \to \pi^+\pi^-\jpsi)
=(0.34\pm 0.14 \pm 0.08)\%$ or 
$\Gamma(\pdp \to \pi^+\pi^-\jpsi)=(80 \pm 32 \pm 21)\kev$,
which is to be compared with an upper limit set by CLEO
of ${\cal B}(\pdp \to \pi^+\pi^-\jpsi)<0.26\%$ (90\% CL).
However, this channel, even if contributing of the order of
$100\kev$ to the decay width, will not be able to account for the
discrepancy previously observed. Radiative $\pdp$~decays
are estimated to amount to at most a few hundred keV.
In addition, the question whether or not there are hadronic 
non-$\DDbar$ decays of 
the $\pdp$ is interesting in the context of mixing scenarios. If
mixing is at work, the modes expected from $\jpsi$ that seem
suppressed at the $\pp$ can give rise to a partial width at the $\pdp$.
An improved understanding of $\pp$ decays will aid in settling this 
question.

\subsection{Decay into lepton pairs\cite{bmumu}}

Studying bottomonium decay into lepton pairs provides access to 
the total width, which at some $10\kev$ for the narrow 
$\Upsilon(1,2,3S)$ resonances is below the typical beam energy
spread of an $e^+e^-$~collider of a few $\mev$, through
$\Gamma_{tot}= \Gamma_{\ell\ell}/{\cal B}_{\ell\ell}$.
In practice, the most precise measurement comes from
employing lepton universality and using $\Gamma_{ee}$
together with ${\cal B}_{\mu\mu}$. 
Measurements of dilepton branching
fractions are interesting in their own right to confront
LQCD predictions (the precision of which has reached the
percent level now), to test lepton universality, and to
compare $\Gamma_{\ell\ell}$ with the hadronic widths
$\Gamma_{ggg, \gamma gg, \qqbar}$.

CLEO studied $\Upsilon(1/2/3S) \to \mu^+\mu^-$ production 
using $1.1 / 1.2 / 1.2 \invfb$ on-resonance and
$0.19 / 0.44 / 0.16 \invfb$ off-resonance data. 
The CLEO results, corrected for interference
with continuum, are:
${\cal B}(\Upsilon(1/2/3S)\to \mu^+\mu^-)^{CLEO}= 
(2.49 \pm 0.02 \pm 0.07) 
/ (2.03 \pm 0.03 \pm 0.08) / ( 2.39 \pm 0.07 \pm 0.10) \%$,
to be compared with the PDG values of\cite{pdg2004}
$(2.48 \pm 0.06) / (1.31 \pm 0.21) / ( 1.81 \pm 0.17) \%$.
This illustrates that the desired precision to keep up with
progress in Lattice QCD has been reached.
Since the CLEO~${\cal B}(\Upsilon(2,3S)$ are found to be substantially
higher, thereby reducing the total width by the same percentage,
predictions for cascade decays such as $\Upsilon(3S)\to \gamma
\chi_{bJ} \to \gamma \gamma \Upsilon(2S)$ are bound to change.

\subsection{Baryon pair production in 
$\jpsi$ and $\chi_{cJ}$ decays\cite{baryonpairs}}

BES used their 58M $\jpsi$ sample to measure ${\cal B}(\jpsi \to \ppbar)
= (2.26 \pm 0.01 \pm 0.14)$.
This is the single most precise measurement of
this branching fraction to date. 
The angular distribution is fit with
the expression $dN / d \cos \theta_p = 1 + \alpha_p \cos^2 \theta_p$,
where $\theta_p$ is the angle between the proton and the beam direction.
Neglecting baryon and quark masses one would expect $\alpha=1$ for all
baryons; including masses yields $\alpha_p=0.66$ and 
$\alpha_\Lambda = 0.51$.
The experimental results are 
$\alpha_p^{exp}=0.676\pm 0.036 \pm 0.042$
and $\alpha_\Lambda^{exp} = 
0.52 \pm 0.33 \pm 0.13$,
in agreement with the prediction.
Proton pairs are produced about twice as copiously
in $\jpsi$ decays as $\Lambda \bar \Lambda$ pairs.
In $\pp$ decays, their branching fractions are comparable.    

Baryon pairs from $\chicj$ decay can be observed through
$\pp \to \gamma \chicj \to \gamma B\bar B$ and compared with
the Color Octet Model prediction that one should expect half as many
$\Lambda \bar\Lambda$ events as $\ppbar$ events. 
These have been made based on $\chicj\to\ppbar$
measurements, which they describe well,
and then generalized to other baryons.  
The BES $\chi_{cJ} \to \Lambda \bar \Lambda$
results
from 14M $\pp$ decays indicated an excess over
this prediction by about a factor of two rather than
a suppression, which has been confirmed as the branching
fractions ${\cal B}(\chicj \to \ppbar)$ have been 
remeasured.

\subsection{``Heavy to Heavy'': Charmonium in $\Upsilon(1S)$ Decays\cite{cleo:bbbartoccbar}}
The Color Octet Mechanism (COM) $\bbbar \to g\ccbar,\ gg\ccbar$ was employed to 
explain $\jpsi$ production rates that could not be attributed to 
the thus far successful Color Singlet Model (CSM), which employs $\bbbar
\to gg\ccbar\ccbar$. 
The two approaches predict different $\jpsi$
momentum spectra as well as angular distributions and branching
fractions for $\Upsilon(1S)\to \jpsi X$. 
A portion of the observed
$\jpsi$ signal will be from $\Upsilon(1S) \to \pp, \chi_{cJ} + X_1
\to \jpsi + X_2$ (not observed before).
The magnitude of this feed-down contribution is also
predicted by the two models. 
 
A CLEO study of charmonium production in $\Upsilon(1S)$ data
intends to
shed additional light onto the question which mechanism is at work. 
Data taken on or near the $\Upsilon(4S)$ resonance is appropriately
scaled and used to calculate the continuum background, which is small
in comparison with the signal.

The inclusive
branching fraction $\Upsilon(1S) \to \jpsi + X$
is measured to be $(6.4 \pm 0.4 \pm 0.6)\times 10^{-4}$, in compliance
with both COM and CSM predictions, both at about $6\times 10^{-4}$.
The process $\Upsilon \to \gamma^* \to q \bar q \to \jpsi + X$ is 
linked with the
continuum process $e^+e^- \to \gamma^* \to q \bar q \to \jpsi + X$ 
and can thus
be estimated relative to the process $\Upsilon \to ggg,
gg\gamma \to \jpsi + X$. The sum of the gluonic reactions 
dominates at a ratio of about 9:1. 
Also, the $\jpsi$ momentum spectrum, scaled according to
$\jpsi$ momentum $x=p_{\jpsi}/p_{max}$ to eliminate beam energy
dependence,
has been measured. The COM predicts a peak at the highest $x$~values,
whereas the CSM shows an accumulation around $x=0.5$.
The measured spectrum peaks at $x=0.3$.
The situation is complicated by the fact that final state 
interactions, which could in principle soften the predicted spectra somewhat, 
have not been taken into account in the predictions. 

Other results of this work include the first determination of 
the branching fractions of and feed-down from $\Upsilon(1S) \to
\pp, \chi_{cJ} + X, J=1,2$, which is found to be a factor of two
above both the CSM and the COM predictions. 
(Since ${\cal B}(\chi_{c0} \to \jpsi \gamma)$ is an order of
magnitude smaller than ${\cal B}(\chi_{c[1,2]} \to \jpsi \gamma)$,
the absence of a signal for $\chi_{c0}$ is not surprising.)

\subsection{``Heavy To Light'' Charmonium Decays\cite{heavytolightdecays}}

Decays of charmonia into light hadrons have often be studied in the light
of the ``12\% rule''. This is a scaling prescription connecting $\pp$ and
$\jpsi$ decays into hadronic final states. 
It allows one to compare the branching fraction ratio with that for 
decay into lepton pairs, which is measured to be $12\%$\cite{pdg2004}. 
Modifications
to this simple picture arise from non-relativistic corrections, form factor
dependence on the two different center-of-mass energies, powers of
$\alpha_s(m_{\pp})/\alpha_s(m_{\jpsi})$, and many more.
Exact agreement with the prediction is therefore not to be expected.
However, even with a more generous view some modes exhibit a substantial
suppression, such as $\rho\pi$ and $K^* K$. 
It has been
conjectured that the suppression is related to quantum numbers and
that vector pseudoscalar final states might be especially affected. Also, 
interference with continuum could play an important role 
as for tiny branching fractions
the resonant and non-resonant cross section may be of comparable magnitude.
Finally, it is possible that the prescription only holds for electromagnetic
processes ($c \bar c \to \gamma^* \to q \bar q$), but not for those
mediated by decay into gluons. This would imply that isospin violating
modes, where the otherwise dominant gluonic process is absent, 
are of special importance to study.
A consistent picture has thus
far not emerged, partly due to lack of experimental data. 

CLEO and BES have brought forward new $\pp \to VP$ measurements.
The most prominent channel is $\pp \to \rho\pi$, which
constitutes a big branching fraction on the $\jpsi$. 
It is not understood why
it is so rare in $\pp$ decays. Another interesting feature
is the different population of the Dalitz plane
from what is seen in continuum and $\jpsi$.
These two show clear $\rho$
bands over some non-resonant background, whereas $\pp$ decays appear to
proceed dominantly non-resonantly. 
To determine what mechanism is at work, a partial wave analysis would be
helpful, which is not possible with the data at hand.

\section{Summary and Acknowledgements}
 
Experimental progress continues in the area of
heavy quarkonia, thereby adding puzzle pieces to our understanding
of many aspects of QCD. It is to be hoped that with future larger
data samples 
more precision studies become feasible and that the remaining
undiscovered states disclose themselves. 

The author wishes to thank her many colleagues 
who provided analysis results, discussion, and guidance.

\end{document}